\documentclass[compsoc,conference,a4paper,10pt,times]{IEEEtran}
\IEEEoverridecommandlockouts
\usepackage{cite}
\usepackage{amsmath,amssymb,amsfonts}
\usepackage{algorithmic}
\usepackage{graphicx}
\usepackage{textcomp}
\usepackage{bmpsize}
\usepackage{xcolor}
\usepackage{lipsum}
\usepackage[colorlinks=true,urlcolor=black]{hyperref}
\def\BibTeX{{\rm B\kern-.05em{\sc i\kern-.025em b}\kern-.08em
    T\kern-.1667em\lower.7ex\hbox{E}\kern-.125emX}}
 
 \usepackage{booktabs}

\usepackage{graphicx}
\usepackage{filecontents}
\usepackage{xcolor}
\usepackage[normalem]{ulem}
\usepackage{enumitem}
\usepackage{multirow}

\newcommand{\myparagraph}[1]{\vspace{4pt} \noindent \textbf{#1.}}

\begin{document}
 
\title{I'm Hearing (Different) Voices: Anonymous Voices to Protect User Privacy}

\author{\IEEEauthorblockN{Henry Turner, Giulio Lovisotto, Simon Eberz, Ivan Martinovic}
\IEEEauthorblockA{\textit{Department of Computer Science,}
\textit{University of Oxford}\\
Oxford, United Kingdom \\
first.last@cs.ox.ac.uk} \\
}

\maketitle

\begin{abstract}
In this paper, we present AltVoice -- a system designed to help user's protect their privacy when using remotely accessed voice services.
The system allows a user to conceal their true voice identity information with no cooperation from the remote voice service: AltVoice re-synthesizes user's spoken audio to sound as if it has been spoken by a different, private identity. 
The system converts audio to its textual representation at its midpoint, and thus removes any linkage between the user's voice and the generated private voices.
We implement AltVoice and we propose six different methods to generate private voice identities, each is based on a user-known secret.
We identify the system's trade-offs, and we investigate them for each of the proposed identity generation methods.
Specifically, we investigate generated voices' diversity, word error rate, perceived speech quality and the success of attackers under privacy compromise and authentication compromise attack scenarios.
Our results show that AltVoice-generated voices are not easily linked to original users, enabling users to protect themselves from voice data leakages and allowing for the revocability of (generated) voice data; akin to using passwords.
However the results also show further work is needed on ensuring that the produced audio is natural, and that identities of private voices are distinct from one another.
We discuss the future steps into improving AltVoice and the new implications that its existence has for the creations of remotely accessed voice services.
\end{abstract}


\maketitle

\section{Introduction}

Systems increasingly utilize remote voice based services for many interactions with users.
These remote systems have also expanded in their capabilities in recent times, now supporting voice authentication to unlock sensitive services (e.g \cite{FirstDirectVoiceID, HSBCVoiceID}), promising smoother and more secure customer interactions by removing the need for users to remember passwords.
Whilst this trend is not a new one, it has been accelerated by the Covid-19 pandemic, with people in many jurisdictions encouraged to avoid in-person interactions.

However, this increased use of voice interfaces comes with significant privacy risks for users. 
Leaks of voice data (e.g., recorded phone calls) not only expose potentially sensitive conversations, but also perpetually discloses the user's voice trait.
Drawing a password-based comparison, while a password can be effortlessly changed after a leak of a password dataset, one's voice trait is unchangeable: once a user's voice is leaked, the secrecy of their voice trait may be compromised forever.
This opens the door for various attacks: both identity linkage and even impersonation may be viable for an adversary who has obtained a sample of a victim's voice.
Furthermore, with the increased prevalence of public videos and audio on social media, the secrecy of one's trait might already be compromised and freely available online.

These concerns are exacerbated by quick advances in voice cloning technology, which shows that it is possible to accurately clone the voice of an individual with just a few seconds of victim audio~\cite{jia2019transfer}.
If the cloned voice is accurate enough, it may be possible to use it directly to impersonate the victim, or to bypass automated voice authentication techniques~\cite{Mukhopadhyay2015, turner2019attacking}.
Furthermore these techniques improve at a fast pace, and the ability to impersonate a victim's voice is only going to become easier.

Recently work has begun on solutions to protect the privacy of users interacting with voice systems.
Several systems have been proposed for service providers, allowing them to protect audio and voice prints they obtain~\cite{Xu2008, Paulini2016, Mtibaa2018, Han2020}.
However, these systems rely on end users adequately protecting their voice information: when such voice information is disclosed or leaked, the systems cannot be safely used any longer.
Furthermore if the user does not trust the provider to protect their voice, then they can not use the system.

More recently, the VoicePrivacy challenge~\cite{VoicePrivacyInitiative} has taken place, in which participants designed systems that would allow a user to anonymize their voice.
The challenge's aim was twofold, to hide speaker identity as much as possible while at the same time limiting the distortion of other speech characteristics to the minimum; part of the goal was to retain as much linguistic content from the original voice.

In this paper, we tackle the challenge of protecting a user's voice secrecy by introducing the AltVoice system, which allows users to replace their own voice with generated unique voices in real-time.
Using AltVoice, users can choose to not re-use their own voice across a multitude of services, but instead they can easily create new unique and secret voices at will, avoiding exposing their original voice with untrusted third-parties.
Differently from previous systems, to create a new voice, AltVoice strips \textit{all} identity information out of the user's voice, this way, it can grant fundamental privacy properties: revocability, unlinkability and noninvertibility.
The proposed system makes use of Speech-to-Text and Text-to-Speech components, which reduce the voice signal to a sequence of words at the intermediate point, resulting in the maximal possible protection of the voice without changing the word content of the speech.
We experimentally validate the performance of our proposed system both in terms of privacy protection and reliability of recognition performance when using authentication systems with a generated voice.
To do this we consider two different attacker models, one aimed at compromising the privacy of a user via an identity linkage attack, and one aimed at impersonating a user who enrolled with a AltVoice-generated voice.
We examine the performance in authentication scenarios using a trained text-independent speaker recognition model based on Generalised End-to-End Loss (GE2E)~\cite{wan2020generalized}.
Finally we examine the effects of differing the algorithm used to create new voice identities, and outline the ways in which this system can be easily upgraded over time, as further improvements are made to speech to text and text to speech systems.
Our results show the system successfully allows a user to assume a new identity, and resists attacks under both of our threat models.
However, the results also reveal that further work is needed in ensuring that generated identities are sufficiently unique and ensuring that the audio generated by the system is both natural sounding and intelligible (i.e. words are not mangled as they pass through the system).

The key \textbf{contributions} of this work are:
\begin{enumerate}
    \item Proposal and implementation of a system (AltVoice) for allowing users to interact with voice processing systems while hiding their voice trait, and without cooperation from the end service provider.
    \item Identification of six different methods to generate private voice identities given the AltVoice architecture, with each identity generation method relying on a user-known secret.
    \item An evaluation of the above system, highlighting its trade-offs and detailing the extent to which the various individual components impact the overall system performance.
\end{enumerate}

\section{Background}
In this section, we provide an overview of typical remote voice processing systems, as well as an overview of the elements used in our proposed system.

\subsection{Remote Voice Processing Systems}
Remote voice processing systems (VPS) have been common for a number of years, and encompass a large variety of services.
Typically the remote aspect of the services means that the interaction occurs through a telephone call, although recently other mediums, such as VoIP programs may be used instead.

The most simple of remote systems are phone calls to call centers, such as those used to provide customer service.
Often these calls are recorded, and there is usually an operator listening at the other end (or a machine processing inputs).
Thus even the least technologically advanced of these systems have privacy concerns for a user, as their voice trait may be recorded and subsequently leaked.
Furthermore users of phones may be unwitting participants in remote voice systems that perform surveillance on behalf of governments or other state entities~\cite{NSASpying}.

These remote systems can also be augmented with additional features.
For example, the phone operator may have software that automatically transcribes phone calls, either at the time or at a later date.
Increasingly extra features are included for voice trait based authentication or identification, to avoid the need for users to answer cumbersome questions to confirm who they are.
For example, in the United Kingdom major banks such as HSBC~\cite{HSBCVoiceID} and FirstDirect~\cite{FirstDirectVoiceID} use voice biometrics as part of their telephone banking service.

Our proposed system should ideally be useable for all types of remote system, including those with authentication.
This requires that any private (or anonymous) voices used by our system can also be used for authentication.
If this is not the case the system will still have utility, and protect the privacy of users utilising remote speech systems without voice based authentication.

\subsection{Speaker Authentication Systems}
Speaker authentication and identification systems can be split into two categories: \textit{text-independent} (TI) and \textit{text-dependent} (TD).
\textit{Text-independent} systems operate on any utterance, where as \textit{text-dependent} systems require the same utterance to be spoken at enrolment and verification time.
Usually this utterance is fixed for all users.

It is also important to distinguish between speaker authentication and identification.
In authentication systems the user claims to be a specific member, and then provides speech data which is used to prove the truth of the claim, to some degree of certainty.
In an identification system an attempt is made to identify who speaks an utterance, but there is no claim of who is speaking beforehand (although a set of candidates may be provided for closed set identification), and we do not necessarily have a confidence threshold that must be met.
In this paper we mostly focus on speaker authentication.

State of the art systems for speaker authentication are based on Deep Neural Networks (DNNs)~\cite{Snyder2018, wan2020generalized}, which take an audio utterance and produce an embedding vector from it, representing the speaker's identity.
As such, the distance between vectors, usually measured by cosine distance, can be used to determine if two samples are spoken by the same user.
State of the art systems achieve Equal Error Rates as low as 3.3\% for TI and 2.38\% for TD~\cite{wan2020generalized}.

A full system also features an enrollment stage, in which several samples are fed to the system, and the embeddings used to create a template for the user, usually by taking the mean embedding.
Later, at verification time, the computed embedding from the sample utterance is compared to the claimed template.
If the distance between template and sample embedding is below a pre-determined threshold it is accepted as spoken by that user, otherwise it is rejected.

Speaker identification systems also use these embeddings, determining the speaker to be the nearest template to a given utterance.

\subsection{Speech-to-Text Systems}
Speech-to-Text (STT) systems produce a transcription of audio input to them.
Traditionally these systems have been built with large amounts of specialist knowledge, but more recently deep learning approaches have been applied to the problem, bringing performance improvements with it.

State of the art systems, such as Deepspeech~\cite{Hannun2014}, use DNNs, specifically Recurrent Neural Networks, to turn spoken audio into character level transcriptions.
These DNNs are trained in an end to end fashion, allowing them to use large datasets and become increasingly robust to variations between speakers, background noise, and other artifacts that may be present in the audio.

Deepspeech (and other similar networks) produce character level transcriptions as their direct output.
A language model is applied to these character level transcriptions to fix errors and ensure valid words are presented, as opposed to phonetic spellings of words.

The accuracy of STT systems is typically calculated using the Word Error Rate (WER), which calculates the number of errors that occur when comparing the transcription to reference the text.
State of the art STT systems achieve WER rates below 10\%, with Microsoft's speech recognition system achieving rates of 5.1\%~\cite{xiong2017microsoft}, a competing system by Google achieving 5.6\%~\cite{chiu2018stateoftheart} and the latest release version of Deepspeech achieving 7.06\%~\cite{deepspeechrelease}.

\subsection{Text-to-Speech Systems}
Text-to-Speech (TTS) systems transform text into audio that appears to be spoken by a real voice.
Most work has historically focused on single speaker TTS, where a large dataset from a single speaker is used to create a model for synthesizing any chosen text.

Recently there has been increased focus on multi-speaker TTS, where a system can synthesise audio from either one of many speakers, or any arbitrary speaker.
These systems consist of a synthesizer and vocoder.
The synthesizer takes an input sequence of phonemes and an embedding that represents the target user, and outputs a set of log-mel spectograms.
The vocoder then takes these spectograms and produces a waveform from this.

The quality of the final output voice depends on several factors, including the quality of the speaker encoder that is used to generate embeddings at training, the quality of the vocoder and the quality of the synthesis network.
Mean Opinion Scores (MOS) are used to assess the quality of the produced audio, where listening tests are performed by humans who give a score for each audio file.

State of the art approaches to TTS achieve MOS of 4.22 for voice in the training data, compared to 4.67 for ground truth audio~\cite{jia2019transfer}.
On speakers unseen in training the MOS is typically worse.

If a TTS system is used to create completely artificial voices i.e never seen, then identity generation is typically performed by random sampling between 0 and 1 and then normalizing the resulting embedding.
Approaches to Voice Privacy using other techniques that create synthetic embeddings have also proposed averaging embeddings from a pool~\cite{Fang2019, srivastava2020design}  and sampling from Gaussian Mixture Models to generate realistic looking embeddings~\cite{turner2020speaker}.
We examine how these embedding selection techniques can be generalized for use with a multi-speaker TTS system later in section~\ref{sec:voiceidgen}.

\section{System Requirements}\label{sec:requirements}

Here we describe the scenario, give a brief outline of how AltVoice works, and present the threat model.

\begin{figure*}[t]
    \centering
    \includegraphics[width=.8\textwidth]{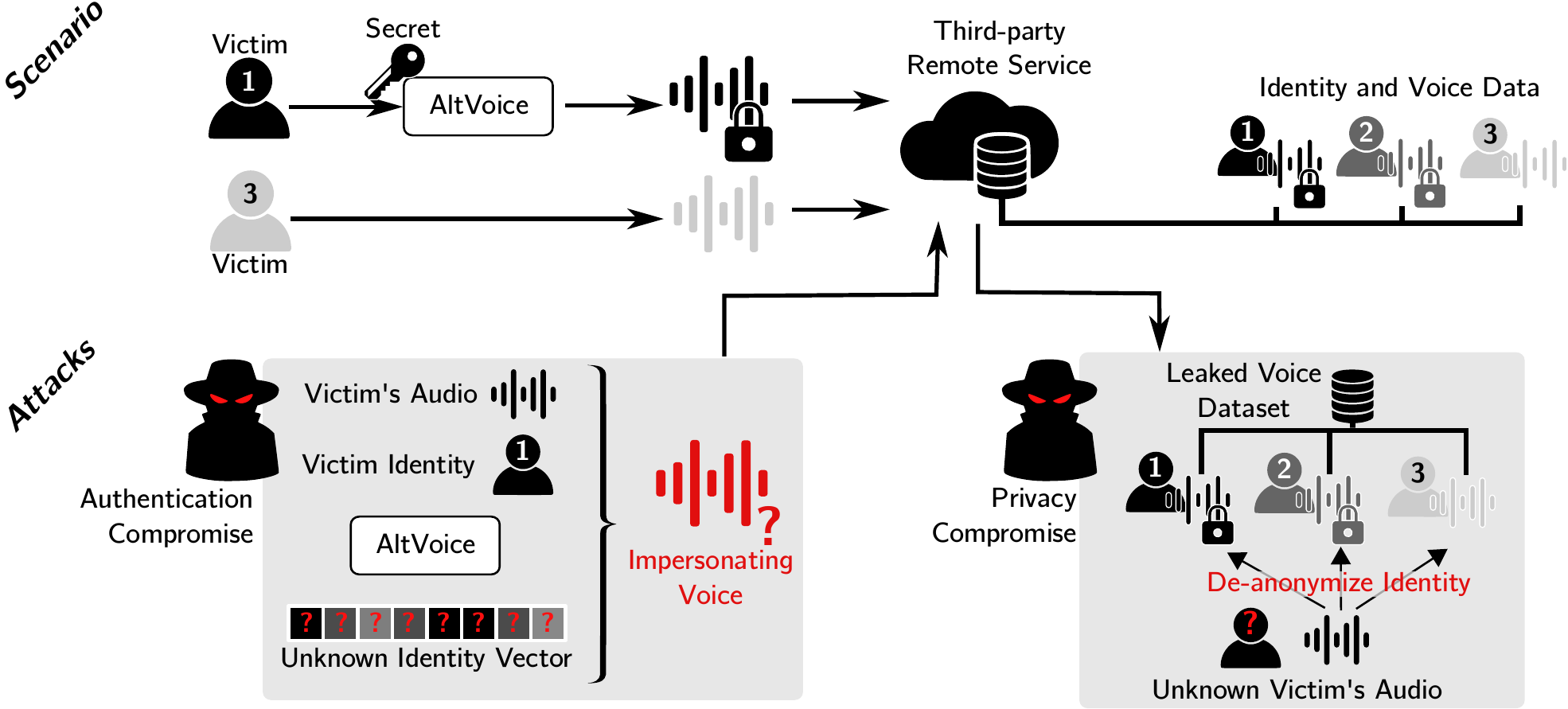}
    \caption{Scenario (top) and attacks (bottom). In the scenario, two users (victims) enrol into a remote voice processing service, which uses and stores users' voices and identity information. The first victim uses AltVoice with a secret to generate a private voice, the second victim uses their original voice. Two different attackers obtain a user's original voice, and attempt two different attacks, either impersonating the victim or de-anonymizing their identity. Using AltVoice protects the first victim from the two attackers.}
    \label{fig:ThreatModel}
\end{figure*}

\subsection{Scenario}

A user of a remote VPS that uses and stores the user's voice data wants to protect the privacy of their voice trait.
Specifically, users are concerned that an adversary could obtain their voice information, either by obtaining access to the remote system data or obtaining audio recordings of the user's voice elsewhere (e.g., audio from social media).
Armed with the user's voice information, an adversary might be able to impersonate the user while interacting with the remote voice system, or they might be able to infer the user's identity by cross-referencing information from other sources.
A representation of this scenario can be seen in Figure~\ref{fig:ThreatModel}.
Users would prefer to be able to use the voice-features provided by the VPS without having to disclose their unique own voice trait in the interactions.

\subsection{AltVoice Design Goals}

\myparagraph{Requirements}
To address the privacy concerns described in the previous section, a system has to \textit{transform} user utterances in a way that \textit{conceals} users' voice identity information, by substituting it with a generated identity.
More specifically, a system must fulfil the following requirements:
\begin{itemize}[leftmargin=0.5cm, topsep=3pt]
    \item\textbf{R1:} Given a user utterance, the system produces a different utterance with identical word content, but different voice identity information. 
    \item \textbf{R2:} The system does not require cooperation from third-party VPS that use voice information, only interaction with the end-user is required.
    \item \textbf{R3:} The identities of the system's generated voices are reproducible: given the same seed or secret, the system generates utterances with the same voice identity.
    \item \textbf{R4:} The system's produced utterances can not be linked to the original user's voice and vice-versa by examining the produced audio.
    \item \textbf{R5:} The diversity among system-generated voices resembles the diversity among natural voices. If diversity is not retained use-cases that require voice-uniqueness (e.g., recognition) may not have the same performance.
\end{itemize}

\myparagraph{AltVoice Outline}
In this paper we present AltVoice, a system that fulfills the above requirements.
AltVoice strips voices of their identity information by transcribing an utterance into words and then synthesizing the words into a new utterance emptied of the users' true voice trait.
AltVoice generates fake voices based on a \textit{user-known secret}, each secret corresponds to a fake voice identity.
In Section~\ref{sec:system_design}, we argue how the system fulfills \textbf{R2} and \textbf{R3} by design.
We use experiments with attack scenarios to validate \textbf{R4} and \textbf{R5} in Section~\ref{sec:exp_diversity} and~\ref{sec:exp_attacks}.
We also use automated speech-to-text transcribers and mean opinion scores to validate how AltVoice retains individual words present in utterances (\textbf{R1}) in Section~\ref{sec:mosexp} and~\ref{sec:werexp}.

\subsection{Threat Model}
In the threat scenario, a victim enrols into a VPS with a AltVoice-generated voice. 
We consider two attacks: (i) Voice de-anonimization attacks and (ii) Voice impersonation attacks; a depiction of these attackers is in Figure~\ref{fig:ThreatModel}.
We first introduce general attacker knowledge and capabilities and then detail each attacker.

\myparagraph{General Attacker Capabilities}
The adversary does not know the secret used by the victim with AltVoice to generate the VPS-enrolled voice. Nevertheless, adversaries have unlimited resources otherwise, specifically they have:
\begin{itemize}[leftmargin=0.5cm, topsep=3pt]
    \item Unlimited amounts of victim's original audio.
    \item A copy of the anonymization system (AltVoice), they can use the system to generate voices.
    \item State-of-the-art voice identification system, which they can use to determine whether two utterances belong to the same individual.
    \item Knowledge that that the victim used AltVoice to enrol with the VPS (rather than the victim's original voice).
\end{itemize}

\myparagraph{Privacy Attacker}
In the privacy compromise attack, the attacker has obtained some of the victim's audio but they do not know the victim's identity. 
The attacker's goal is to de-anonymize the victim by cross-referencing the victim's audio against a voice dataset that contains links between users' identities and natural voices.
In practice, the attacker checks whether the victim's audio matches voices in the voice dataset, when they find a match, they infer the victim's identity to be that associated with the matching voice in the dataset.
We assume that the victim is in the voice dataset that the attacker has obtained.

We assume that the privacy attacker uses the voice data they have obtained to perform their attack, and does not attempt to use other information, such as word choices or sentence lengths to infer the identity, and deem such methods out of scope. 
Whilst stylometric approaches that use this information may help an attacker, it is hard to model them effectively, and our proposed system is not designed to defend against them.
Future work could examine methods that change the word content of utterances to defend against this style of attack.

\myparagraph{Authentication System Attacker}
In the authentication attack, the attacker has obtained the victim's audio and they know the victim's identity.
The attacker's goal is to impersonate the user when interacting with a VPS which uses speaker recognition (where the victim is enrolled).
In practice, the attacker can use AltVoice to generate a private voice utterance and attempt to log in to the VPS with the generated utterance.\footnote{For this work, we consider text-independent speaker recognition is in place. We discuss text-dependency in Section~\ref{sec:limitations}.}
Note that while we evaluate impersonation considering a single attacker's attempt, in practice a VPS typically allows multiple attempts before throttling, with the number of attempts depending on the VPS access control policies and is outside the scope of this paper.

\section{System Design}\label{sec:system_design}

\begin{figure*}[t]
    \centering
    \includegraphics[width=.85\textwidth]{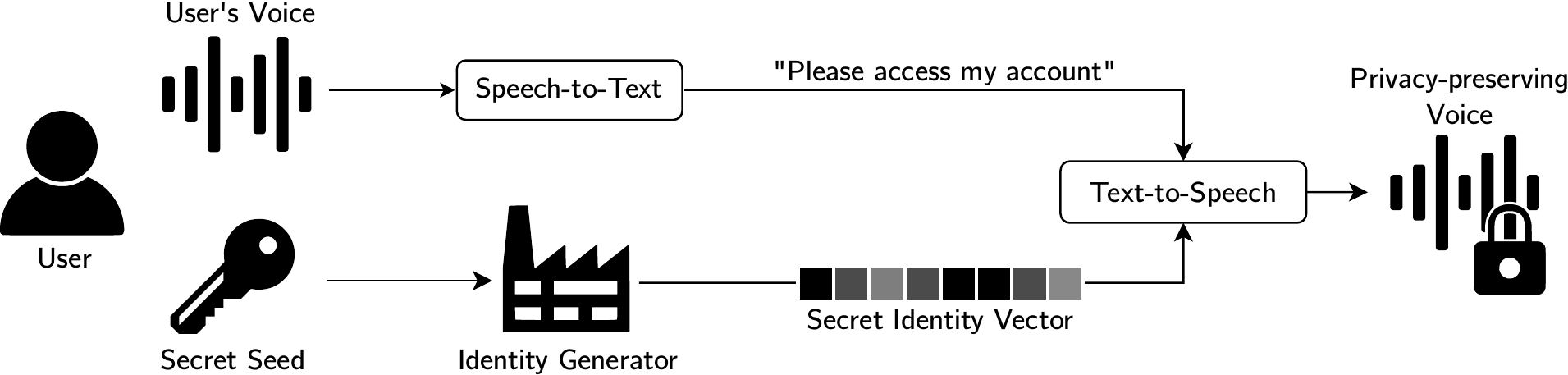}
    \caption{AltVoice system diagram showing the key components of the system and how they interoperate with one another. AltVoice uses the user's voice and a user defined secret to generate a private voice which can be used with remote VPS.}
    \label{fig:SysDiag}
\end{figure*}

In order to meet the requirement set out in section~\ref{sec:requirements} we design the AltVoice system.
The system aims to produce audio, conveying the same words as that being spoken by a user, but that appear (both to a speaker identification system and to a listener) to be spoken by a different individual, in as close to real time as possible.
In order to prevent any linkage between the initial identity and outcome identity of the audio, we design our system so that only a textual representation of the audio is present at the midpoint of the audio generation process.

We do this through sequential application of Speech-To-Text (STT) and Text-To-Speech (TTS) systems, converting the speech of the user into the text content of their speech, followed by turning this text back into speech spoken by a different speaker.
The identity of this new speaker is provided by an identity generation system, and we evaluate several candidate generation methods in this work.
By reducing all the information from the original speech signal to just its text content at the midpoint of the system pipeline, we removal the maximal amount of speech information present, thus giving the strongest privacy guarantees possible short of changing the words spoken.
An overview of the system can be seen in Figure \ref{fig:SysDiag}.

In the remainder of this section we discuss the specific sub-systems we implemented to fulfil each part of the system in further detail.

\subsection{Implementation}
The AltVoice system is made up of several sub components.
In this section we discuss the considerations that need to be made for each specific component, as well as detailing the specific implementations we selected for this version of AltVoice.

\subsubsection{Speech to Text}
The STT system forms the first part of the pipeline, and its accuracy is critical for ensuring that the final spoken audio features the correct words.
Errors that occur in this section will be propagated (and potentially worsened) through the rest of the model.

Another important consideration for the chosen model is its execution speed - if it is too slow then the lag between speech and transformed audio being produced will be increased.
This is further impacted by the utterance length required to produce a transcription (some models require full sentences).

In our implementation of AltVoice, we utilize the Deepspeech system~\cite{hannun2014deep}, which is freely available with pre-trained models~\cite{deepspeechrelease}.
Deepspeech has state of the art performance, both in terms of word error rate (WER) (Down to 7.06\% on a benchmark audio set) and execution speed.

The Deepspeech system uses an RNN model to transform audio into a sequence of character-level transcriptions, which in turn are processed by a language model to help fix errors such as phonetic misspellings of words.

Customizing the language model of Deepspeech can improve accuracy on domain specific tasks.
In this example we evaluate AltVoice as a general purpose framework, and as such use the default language model, however in practical usage of AltVoice it may be beneficial to swap the language model to a domain specific one.

We augment the execution of the Deepspeech model by using voice activity detection (VAD) to detect pauses in speech, cutting the speech input and passing it to Deepspeech as often as possible.
In doing so we can reduce the total system lag on producing output audio, by not requiring the system to wait for another indicator that speech is complete.
There is a potential trade off here, in that overaggressive VAD may impact the overall WER of the system.
In our implementation of AltVoice we use the publicaly available WebRTC~\cite{webrtc} VAD.

\subsubsection{Text to Speech}
The Text to Speech (TTS) system turns the text, and a speaker embedding, representing the new identity of the voice, into audio data.

Our overall multi-speaker TTS model is based on that of \cite{jia2019transfer}, which utilises a speaker encoder network to provide an embedding of a voice to the synthesizer, along with the text to be produced.
The sequence of mel spectrograms outputted by this is then passed to a vocoder, which produces output audio

In our implementation we use the open source Mozilla TTS implementation of Multi-speaker TTS\footnote{https://github.com/mozilla/TTS}.
From this library we use Tacotron 2~\cite{shen2018natural} for our spectogram predicition network, and use a fullband MelGAN~\cite{yang2020multiband} as our vocoder.
The Tacotron 2 model is trained on the VCTK dataset\cite{Veaux2017CSTRVC}, with the Vocoder train on the LibriTTS dataset~\cite{libritts}.
For training the Tacotron2 model a GE2E model trained on LibriSpeech dataset is used~\cite{librispeech} as the speaker encoder network.

The choice of both of these components has a large impact on both the quality of the audio produced and the diversity of the voices produced.
The Tacotron 2 model achieves MOS that are close to 4.526 vs 4.582 for ground truth speech, demonstrating its ability to create natural sounding speech.
However, recent work has demonstrated that the datasets used in training impact the perceived quality of speech (via MOS testing), as well as the signal rate the speech was recorded at~\cite{cooper2020pretraining}.

We perform MOS tests on our audio in section~\ref{sec:mosexp}, to verify the performance of our system.
This is especially important as previous examination of naturalness for fictitious speakers has been limited, with~\cite{jia2019transfer} demonstrating the technique on a set of 10 speakers generated using the random technique, which we also evaluate.

\subsubsection{Voice Identity Generator}
\label{sec:voiceidgen}
The Voice Identity Generator (VIG) provides the embedding to the TTS system which defines the output voice.
Limited prior work exists on Voice Identity Generation for TTS, with demo examples of multi-speaker TTS just generating vectors using random numbers and then normalizing the output vector.

In the voice privacy challenge several solutions, including the baseline, contained an element for voice identity generation.
The work of Fang et al \cite{Fang2019} examined potential techniques for voice identity generation within their X-vector anonymisation technique, used as the baseline for the voice privacy challenge, based on averaging embeddings from a subset of a pool of users,  either using a random selection or (furthest) distance based measures.
Turner et al. proposed in \cite{turner2020speaker} using a combination of a PCA coupled with a GMM to generate new embeddings which mirror the distribution of those found in the real world.

For our voice identity generator we examine six possible techniques:

\myparagraph{Random Generation} Sample a value for each embedding feature from a normal distribution of mean 0 and standard deviation 1.
The final new identity embedding is then normalized.
This technique is based on that of Jia et al. \cite{jia2019transfer} to create fictitious voices.

\myparagraph{Random Generation in PCA Space}
Conceptually similar to Random Generation, but instead of performing the random sampling in the embedding space it is conducted in a Principal Component Analysis (PCA) space.
We fit the PCA on training data from applying the GE2E extractor used in the training of the TTS system to the VoxCeleb 1 and 2 Development Datasets.
We create one PCA transform for each gender, and fit the PCA so that it captures 95\% of the variance in the data.
For each component of the PCA space, we then determine the mean and standard deviation.
When generating a new identity, we then sample from a normal distribution with the given mean and standard deviation for each component in PCA space, before performing an inverse PCA transformation to give an embedding in the original embedding space.

\myparagraph{Mean Pool Subset} We follow the technique proposed in~\cite{Fang2019} for anonymised voice conversion in the X-vector space. 
We use a pool of all of extracted embeddings of the VoxCeleb 1 and 2 Development sets, and average ten embeddings taken from the pool to give us our final new identity embedding.
We create a separate pool for each gender.

\myparagraph{PCA + Gaussian Mixture Model} We follow the approach proposed in \cite{turner2020speaker, turner2022generating}, which was applied to an X-vector space for anonymisation with voice conversion.
As proposed in this work, we use a separate GMM for each gender, fitting the PCA on 95\% of the variance, and using 20 components for the Gaussian Mixture Models fitted in the PCA space.
We use the VoxCeleb 1 and 2 Development sets for training the PCA and GMM.

We evaluate each of these four techniques when examining the diversity of voices created by the AltVoice system, as it is the generation strategy, coupled with any biases introduced by the TTS which produce the overall voice identity, and will be responsible for the diversity of the produce private voices.

\myparagraph{Pool Selection}
Using the pool of embeddings extracted from the VoxCeleb 1 and 2 Development sets, we select a single identity to use as the new anonymous identity for the user.
In an optimal TTS system, this would be the same as cloning an existing voice from the pool.
We use a different pool for each gender.

\myparagraph{Training Selection}
This is similar in concept to the pool section method, but the embedding is taken from the set of user embeddings used in training the TTS model.
As the model has been trained on these embeddings, we would expect the TTS system to be able to produce higher quality audio for these embeddings.

\subsection{Usage in Practice}
In practice AltVoice would be implemented as a `Dialer' application that allows a user to make phone calls. 

For all of the identity generation methods a secret user-known seed is used to derive the embedding.
This seed is used to instantiate a random number generator, which is then used either directly, for sampling from models, or from selecting from sets of existing users.
The application could manage the storage of these seeds (and thus the associated voices).
By default it would generate a new identity for each different service that is contacted, but it would also be possible to configure the application to change identity when required by the user, and thus change it for different contexts with the same phone number, or to maintain the same identity across multiple calls to different service providers.
In cases where multiple voices are used for a single service, it would be necessary for the user to press a button to change the voice.

Finally in practice a user may not wish to provide input to the system as voice, and could instead provide text input direct to the TTS component of the system.
This would reduce the complexity of the system, and yield higher accuracy as mistakes could no longer occur in the STT component.

\begin{figure*}[t]
    \centering
    \includegraphics[width=0.98\textwidth]{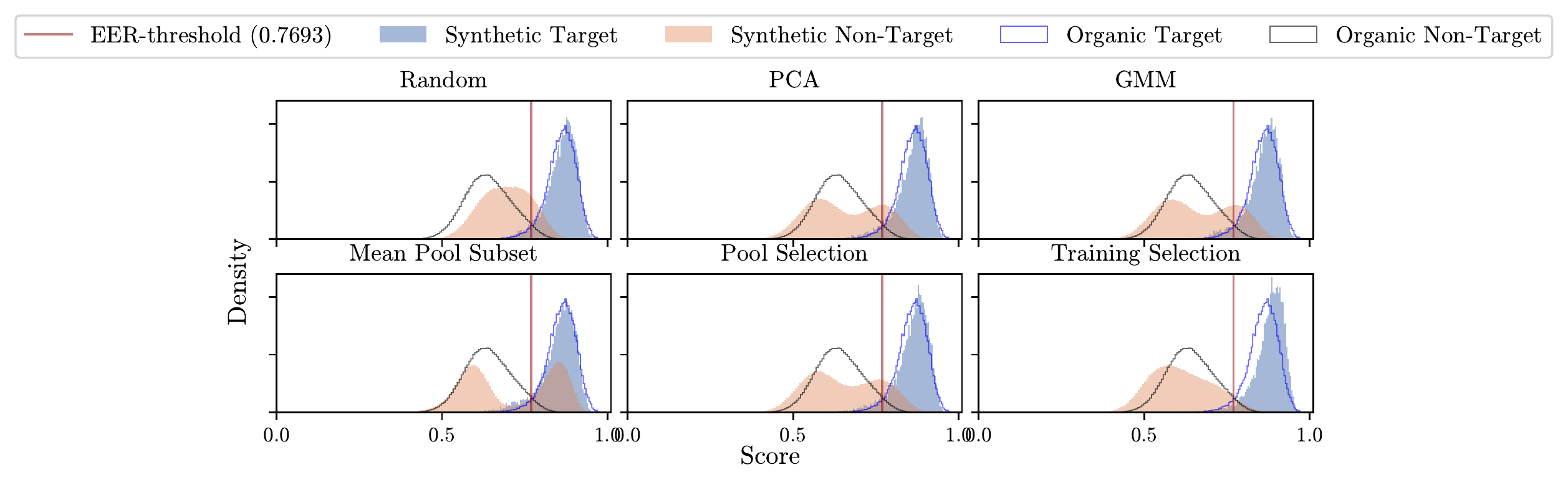}
    \caption{Distribution of pairwise cosine similarity scores between naturally occurring voices (organic) and voices produced by AltVoice for each of the identity generation methods. Target means that the pair compares a identity template with an utterance belonging to the template owner, non-target means the opposite. The EER-threshold is computed using naturally occurring voices.}
    \label{fig:CosPopln}
\end{figure*}

\section{Experimental Evaluation and Results}\label{sec:experiments}

In this section we evaluate AltVoice, focusing on the requirements specified in Section~\ref{sec:requirements}. 
We first analyze the diversity of anonymized voices, followed by the resilience of AltVoice to attacks, and then look at how AltVoice-generated utterances maintain word content.

\subsection{Diversity of Anonymized Voices}\label{sec:exp_diversity}

\begin{figure*}[t]
    \centering
    \includegraphics[width=0.8\textwidth]{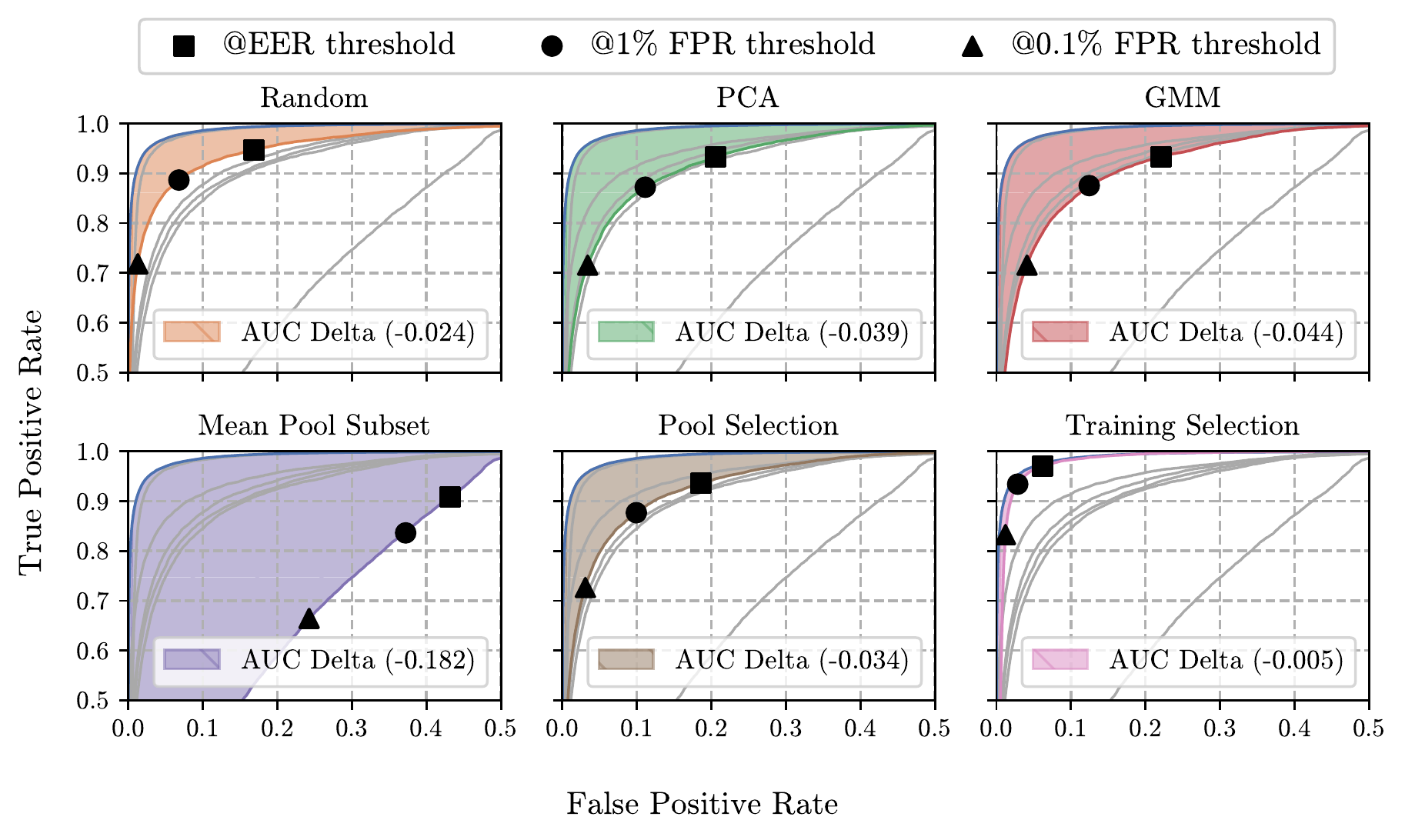}
    \caption{ROC Curves obtained by comparing target and non-target vector identity pairs, for each identity generation method. The solid blue line reports the performance among naturally occurring voices, which leads to an equal error rate (EER) of 3.84\%. Markers on the curves report the performance trade-off for various ways of selecting a test-time recognition threshold based on the performance among natural voices: either at EER level, at 1\% FPR or at 0.1\% FPR. Shaded areas report the method AUC difference compared to the AUC obtained using natural voices. In this chart a true positive is two audio samples with the same target identity matching, whereas a false positive is two samples with different identities matching.}
    \label{fig:diversity_ROC}
\end{figure*}
\label{sec:diversity}

The diversity of anonymized voices is crucial when considering the authentication system compromise threat model: too little diversity and users will have less security than normal voices as attackers would be able to authenticate with a random voice.
Poor diversity will lead to worse performance in an authentication system, failing to meet R5.

\myparagraph{Experiment Setup}
To test voice diversity, we use a state-of-the-art  text-independent speaker recognition method based upon the GE2E system proposed in~\cite{wan2020generalized}.
GE2E extracts multi-dimensional identity embeddings from voices: we expect that pairs of identity embeddings obtained from AltVoice-generated voices are as dissimilar as pairs of identity embeddings obtained from natural (or \textit{organic}) voices.
For this analysis we bypass the STT part of AltVoice, and we generate AltVoice voices directly from text files, using sentences from the Mozilla Common Voice dataset in the English language~\cite{mozillacommonvoice}.
For each of the six identity generation methods introduced in Section~\ref{sec:system_design}, we produce 500 voice identities, and generate 30 spoken sentences for each of these.
For each identity, we use the identity embeddings extracted from 10 out of the 30 sentences to enroll an identity template (by taking the average of them), we reserve the other 20 utterances to be used as positive and negative trials.
We compare the identity templates with the remaining 20$\times$500 utterances, by selecting sets of  \textit{target} and \textit{non-target} pairs (i.e., target when the utterance belongs to the same identity as the template, non-target otherwise) and storing their distance scores using cosine similarity~\cite{wan2020generalized}.
Finally, we use the distribution of cosine similarities to compare natural and AltVoice-generated voice diversity.
We use the Voxceleb~1 and~2 test datasets~\cite{nagrani2017voxceleb} for the natural voices.
This dataset contains audio data that is scraped from videos of celebrities speaking, and contains over 1 million utterances from more than 6000 celebrities.

\myparagraph{Results - Similarity Distributions}
Figure~\ref{fig:CosPopln} shows the distributions of distance scores for the six identity generation methods.
The distribution of target pairs i.e., same identity, are similar to those for naturally occurring voices.
We see that these are shifted slightly right than the original audio, an effect particularly noticeable for the training identity generation method, suggesting utterances from the same identity have less variation for AltVoice-generated voices than for naturally occurring voices.

Comparing non-target pairs, we see a disparity between the naturally occurring distribution and those for the identity generation methods.
For the PCA, GMM and Mean Pool Subset models we see two peaks, corresponding to each gender.
This suggests that these methods create voices for each of the genders that are more similar to one another than naturally occurring voices of that gender.
For the Random method we see the distribution is shifted to the right, implying that the voices are more similar than naturally occurring ones.

The pool selection method is dual peaked, implying that the underlying TTS network is responsible for some of the increased similarity between voices of the same gender, as opposed to this being a consequence of just the identity generation method.
The training selection method shows a peak to the left of the baseline, but with a wider distribution.
This suggests some overfitting may be occurring, due the leftward shift of the peak.
Similarly it may be that voices of the same gender are more similar, causing the long tailed distribution.

\myparagraph{Results - ROC curves}
Figure~\ref{fig:diversity_ROC} report receiver operating characteristic curves obtained with the pairwise similarity scores.
Figure~\ref{fig:diversity_ROC} shows that the performance for each of the methods i.e., the area under the curve (AUC), is worse than the AUC computed among naturally occurring voices (blue solid line in Figure~\ref{fig:diversity_ROC}); the gap between the natural AUC and the AltVoice-generated AUC is reported in each plot for each method.
The voices in Training methods performs best, with a small difference of 0.005 AUC from the result obtained with naturally occurring voices.
It should be noted however that we are only limited to 96 voices used in training, so in this case there is a 1/96 chance a voice generated by a user would be the same as another voice generated by another use.

\subsection{Threat Model Attacks}\label{sec:exp_attacks}
Here, we study the performance of AltVoice against two attacks: a privacy compromise attack, and an authentication compromise attack.

\myparagraph{Audio Used}
To evaluate the attacks, we re-use the same audio generated in the previous section: we generate 500 voice identities for each of our proposed identity generation methods and for each private voice we create 30 spoken utterances.
As before, we use source sentences obtained from the Mozilla Common Voice dataset in the English language.
We use the speakers in the VoxCeleb 1 and 2 test datasets as naturally occurring voices~\cite{nagrani2017voxceleb, chung2018voxceleb2}.

\subsubsection{Privacy Compromise}
\myparagraph{Experiment Setup}
To conduct the privacy compromise attack, the attacker has obtained a sample of a private voice, and wishes to identify the speaker the voice belongs to, matching against known candidate voices.

To simulate the adversary's knowledge of a voice dataset, we  select a set of 20 speakers from VoxCeleb test set, for each of them we compute an identity template available to the adversary.
Then, we use the same 20 speakers with AltVoice, we generate a private voice for each speaker and we produce one utterance for each of them: the adversary needs to match private voices with speaker identities.
To do the matching, the adversary computes the cosine similarity between the utterance embedding, and the embedding of the user template, and select the most likely speaker as the nearest voice.
Cosine similarity is used as the similarity measure when training the GE2E network, and thus is the best method for an attacker to determine the similarity of two embeddings.

We conduct 100 rounds of the experiment for each identity generation method.
Throughout the attack, the adversary only considers voices that are the same gender as the original voice (in the case of the random identity generation method gender is ignored).

\begin{table}[t]
    \centering
    \caption{Privacy attack results. The table shows how the attacker can successfully de-anonymize a non-protected victim (baseline) with high success, while when using AltVoice with the various identity generation methods the attack success rate decreases.}
\begin{tabular}{clrr}
\toprule
 & Generation Method &  Success Rate (\%) &  95\% CI (pm) \\
\midrule
    &        Baseline &            97.585 &        0.002 \\
    \hline
\multirow{6}{*}{\rotatebox[origin=c]{90}{\textbf{AltVoice}}} &             GMM &            10.355 &        0.618 \\
    &          Random &             7.713 &        0.557 \\
    &             PCA &             8.445 &        0.460 \\
    &            Mean Pool Subset &             8.826 &        0.370 \\
    &        Pool Selection &             9.063 &        0.335 \\
    &        Training Selection &             9.276 &        0.310 \\
\bottomrule
\end{tabular}

\label{tab:PrivacyAttack}
\end{table}

\myparagraph{Results}
We report the results of the attack in Table~\ref{tab:PrivacyAttack}.
We find that conducting the privacy compromise attack on normal voices (without the AltVoice protection), the adversary identifies the speaker with a success rate of 97.6\%.
This is as expected, as the encoder network used is highly accurate, and as such it is almost always the case that the nearest speaker to the sample is the correct one.
Applying AltVoice, the attack success rates are reduced significantly.
For the methods with gender, the mean success rate is just below 10\%, which is the expected outcome from guessing randomly out of 20 identities to choose from.
The reason for mean success rate being just below 10\% could be due to the synthetic voices not being as spread as naturally occurring ones, as seen in the experiments in Section~\ref{sec:diversity}, meaning that the same incorrect guesses are made for large parts of the voice spectrum.
The random generation method achieves a 7\% performance, slightly higher than the 5\% from purely random guessing, again likely caused by the random voices having a different distribution than normal voices.

\subsubsection{Authentication Compromise}
\myparagraph{Experiment Setup}
\begin{figure*}[t]
    \centering
    \includegraphics[width=0.7\textwidth]{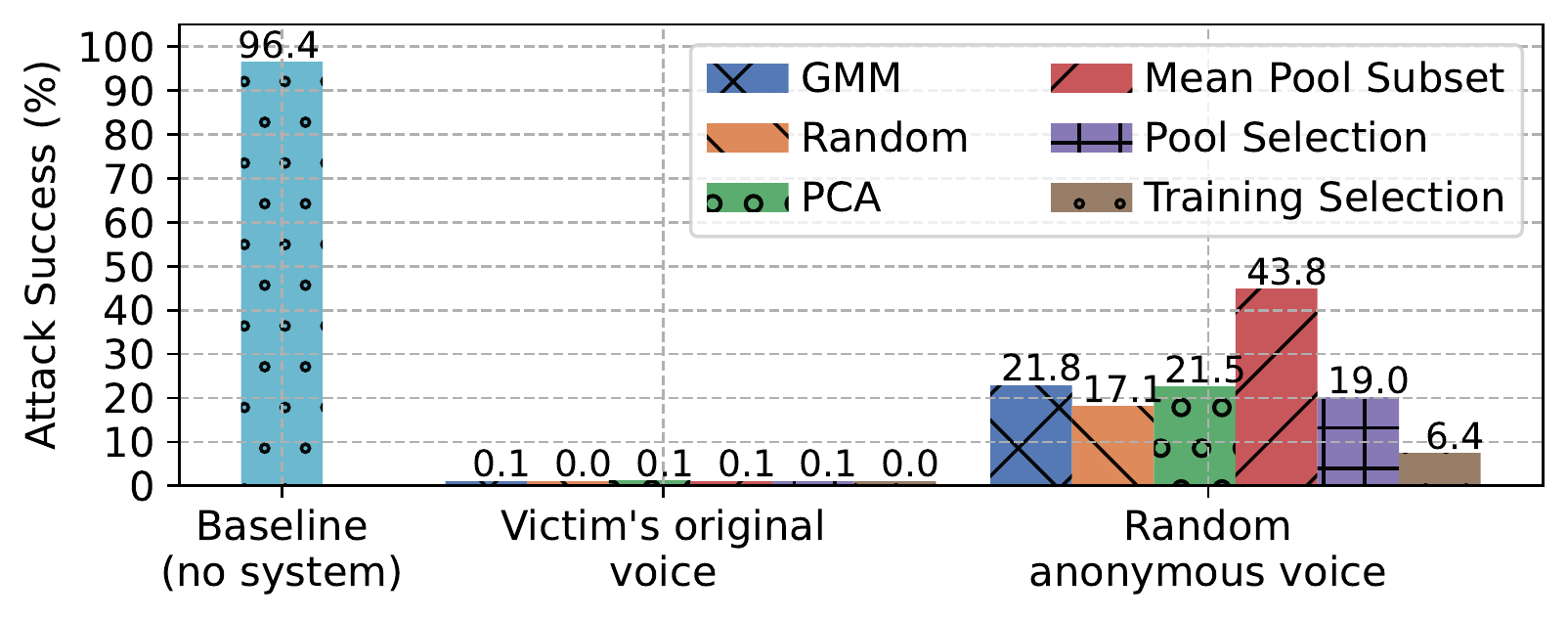}
    \caption{Adversary success rates under our authentication attack. Baseline refers to the attack success rate when no AltVoice is in place (voices are unprotected). Victim's original voice refers to the attack success rate with AltVoice in place and the adversary using the victim's original voice for impersonation. Random anonymous voice refers to the attack success rate with AltVoice in place and the adversary using a randomly generated AltVoice voice for impersonation.}
    \label{fig:AuthAttack}
\end{figure*}
To conduct the authentication compromise setup, we consider two scenarios, one in which the victim has enrolled with a VPS with their natural voice to act as a baseline, and one where they have used a private AltVoice voice.
We re-use the same speaker verification system used in Section~\ref{sec:diversity}.
We follow a similar procedure to evaluate both scenarios:
\begin{enumerate}
    \item Select a victim from the set of possible speakers (Voxceleb 1 and 2 test datasets).
    \item Create the victim's natural template from 10 utterances, or create a private voice for the victim and a template for this with 10 AltVoice-generated utterances.
    \item Adversary presents an utterance in an attempt to access the system, using one of two strategies: (i) Use a natural utterance belonging to the victim or (ii) Use a AltVoice-generated utterance with the same identity generation as the victim (but different secret seed).
\end{enumerate}

We treat a successful attack as one where the similarity score exceeds the threshold value at EER using the Voxceleb 1 and 2 test dataset, i.e., an attack is successful when the adversary successfully impersonates the user.
We conduct 10000 trials for each of the scenarios.

\myparagraph{Results}
Figure~\ref{fig:AuthAttack} shows a comparison of the results for the baseline and both of the attack strategies employed by the adversary.
In the Baseline scenario, we find that an attacker is successful 96.4\% of the time.
This is intuitively the case, as if an attacker has perfect copies of a voice, they should be able to impersonate that user with a success rate the same as the system's true positive rate.

Testing the victim's original voice against a private rate we get a success rate of 0.1\% for all identity generation methods.
This shows that the private voices generated by our system are only related to the original voice by their textual content, and thus the identity embeddings are fully independent.

With an attacker trying random anonymous voice we see varied results across the attack techniques.
The Mean Pool Subset method performs worst, with the attacker selecting a random voice that is closer than the threshold 43.8\% of the time.
Selecting a voice with the Training identity generation performs strongest, with the attacker randomly selecting a colliding voice 6.4\% of the time.

\subsection{Perception of Audio Quality}
\label{sec:mosexp}
While the (perceived) quality of generated audio is not important for automated systems (beyond the WER), it is crucial for communicating with humans.
If the generated voice sounds obviously synthetic, the human may terminate the call or refuse to engage in sensitive transactions (e.g., in phone banking).
Similarly to previous works, we perform Mean Opinion Score (MOS) tests to measure the quality of the voices that are produced by our technique.

\myparagraph{Experimental Setup} 
In order to obtain MOS, we use a crowdsourcing approach on Amazon Mechanical Turk.
To achieve suitable control and comparability to existing work, we follow the recently established ITU-T recommendation P.808~\cite{p808standard}, with each audio file being rate on a scale of 1 (poor) to 5 (excellent).
This recommendation includes multiple techniques to obtain high-quality responses, including a qualification phase, setup phase and training phase. 
During the qualification phase, we verify that workers are native English speakers, use stereo or in-ear headphones (rather than speakers), are of normal hearing and operate in a quiet environment.
In line with P.808 recommendations, we use five audio samples for training that are manually selected to cover a range from the worst to best samples.
These are presented in a random order but are identical for all workers.
During both the training and rating phase, we use gold standard and trapping questions to confirm worker attentiveness.

Following the completion of these phases, we present a total of 700 audio samples -- 100 baseline clean audio files from the LibriSpeech dataset, and the same set of files fed through the AltVoice system with a new private identity randomly generated for each file.
These samples are presented in sets of 12 and workers are compensated for each set they complete.
Each set includes samples from all of the seven conditions (Baseline + 6 generation methods) in order to prevent workers who don't rate all 700 samples from biasing the results.
In line with P.808 recommendations, we incentivize workers to complete at least 50\% of total samples.
Overall, each sample is rated by at least 8 distinct workers.
We received ethical approval from our institution to perform this experiment, reference SSD/CUREC1A\_CS\_C1A\_21\_010.

\begin{table}[t]
    \centering
    \caption{Mean Opinion Scores for the baseline audio and each identity generation scheme. All AltVoice-generated audio performs worse than the Baseline.}
   \begin{tabular}{clrr}
\toprule
 & Identity Generation Method &   MOS &  CI ($\pm$, 95\%) \\
\midrule
 & Baseline & 4.230 &         0.054 \\ \hline
\multirow{6}{*}{\rotatebox[origin=c]{90}{\textbf{AltVoice}}} &  Random & 3.015 &         0.067 \\
 &     GMM & 2.514 &         0.059 \\
 &     PCA & 2.548 &         0.058 \\
 &    Mean Pool Subset & 2.438 &         0.054 \\
 & Pool Selection & 2.495 &         0.052 \\
 & Training Selection & 2.607 &         0.062 \\
\bottomrule
\end{tabular}
    \label{tab:MOSResults}
\end{table}

\myparagraph{Results}
Our MOS results are presented in Table~\ref{tab:MOSResults}.
All AltVoice generated audio achieves worse MOS scores than the Baseline audio from organic speakers.
Of the identity generation techniques Random performs best with an MOS of 3.015.
Our MOS results are in contrast with the higher ($>$4) MOS scores reported in the TTS paper~\cite{jia2019transfer}, suggesting that such systems will need further improvement before they can be integrated out-of-the-box into more complex pipelines such as AltVoice.
Scores of 2.5 on a MOS scale mean that in most cases it would be obvious that the audio is not from a natural (human) speaker.

\subsection{Word Error Rates}\label{sec:werexp}
We evaluate the word error rates of AltVoice when applied in an end-to-end manner, to determine how much word information is lost.
There are two potential points at which errors may be introduced into the system: (i) when initially performing speech to text and (ii) when synthesizing speech from text.
The model used for performing STT has a WER rate of 7.06\% on clean test data, giving this as an empirical upper bound on performance.

\begin{table*}[t]
    \centering
    \caption{ASR Recognition Rates for 1,000 Audio files generated by our system from the LibriSpeech test-clean dataset. Transcriptions are produced with a local version of Deepspeech 0.9.3 and on Google Cloud Speech-to-Text Service in May 2021.}
\begin{tabular}{lllll}
\toprule
System & \multicolumn{2}{l}{Deepspeech} & \multicolumn{2}{l}{Google Cloud} \\
Metric &                   WER [CI] &                   WIL [CI] &                   WER [CI] &                   WIL [CI] \\
\midrule
Baseline &                 9.74\% &                16.82\% &                 9.74\% &                16.82\% \\
Random   &  42.41\% [40.84, 43.98] &  57.54\% [55.84, 59.24] &  47.92\% [46.49, 49.35] &  67.11\% [65.64, 68.59] \\
GMM      &  24.48\% [23.50, 25.47] &  36.65\% [35.38, 37.92] &  30.27\% [29.04, 31.51] &  46.46\% [44.86, 48.05] \\
PCA      &  25.42\% [24.48, 26.37] &  37.88\% [36.68, 39.08] &  31.57\% [30.37, 32.77] &  48.11\% [46.56, 49.66] \\
Mean Pool Subset     &  22.93\% [22.15, 23.70] &  34.72\% [33.70, 35.74] &  28.67\% [27.58, 29.76] &  44.34\% [42.89, 45.80] \\
Pool Selection &  24.33\% [23.42, 25.23] &  36.52\% [35.34, 37.69] &  29.88\% [28.68, 31.08] &  46.01\% [44.41, 47.61] \\
Training Selection &  28.99\% [27.86, 30.12] &  42.47\% [41.11, 43.82] &  35.80\% [34.28, 37.31] &  53.30\% [51.45, 55.14] \\
\bottomrule
\end{tabular}
\label{Table:WER}
\end{table*}

\myparagraph{Experiment Setup}
We apply AltVoice to a set of 25 randomly selected audio files from the LibriSpeech test-clean dataset, transforming each file to the same private identity. 
This is repeated 100 times, and for each of the 6 proposed identity generation methods.

Each generated audio file is transcribed using both the Google Cloud Speech API and Deepspeech.
We also transcribe the original 25 files, to give a baseline for comparison.
We evaluate the differences in this transcribed audio using Word Error Rate (WER) and Word Information Lost (WIL).
Often only WER is used, however WIL is preferred for speech use, as it measures the proportion of word information communicated~\cite{Morris2004}.
Note that by definition WIL can never be higher than WER for a given set of sentences.

\myparagraph{Results}
The results for each of the techniques can be seen in Table~\ref{Table:WER}.
We see that all the identity generation methods have a large degradation from the baseline in both WER and WIL.
This degradation would be discernible to listeners, and could lead to sentences with lost meaning or confusing words embedded in them.

There is some difference in both of the metrics for the various techniques.
Most notably we see that performance is worst for the random identity generation method.
This could be a result of there being some relationships between the values in the identity embedding, which are lost when all values are set randomly, resulting in identities that the system struggles to produce audio for.
Across the other methods we see broadly similar results, with the Mean Pool Subset method performing best.
This could be because averaging many embeddings results in an embedding central in the overall space, and thus fairly neutral and easier to synthesize quality audio for.

Interestingly the Training Selection method performs worse than the Pool Selection method, despite the training identities having been seen before.

\section{Discussion}
In this section we discuss the identity generation performance, system limitations, and the ways in which we expect the individual components to improve over time, highlighting directions for future work.
We also discuss the general implications of the existence of this system, and in particular new considerations for designers of remote voice based systems as a result of the existence of private voices.

\subsection{Identity Generation Performance}
The results from each of the experiments paint a mixed picture for the identity generation techniques we evaluate, with no technique being a strong performer across all categories.

Generally the techniques that create structured (i.e not purely random) synthetic embeddings from existing models performed poorly in the diversity and attack sections, with the homogeneity of the embeddings they create meaning that an adversary can generate a fairly similar voice with few attempts.
The random generation scheme, produces more diverse voices, with better MOS performance, but at the expense of much higher WER and WIL scores.

Both the Pool Selection and Training Selection identity generation methods produce more diverse voices, and perform better under both attack scenarios.
The training selection method in particular performs very strongly under both attack threat models, as well as performing second best for MOS, although has a slightly higher performance cost for WER and WIL than the synthetic techniques.

Going forward further work is needed on generating good synthetic identities.
Alternatively further improvements to the models performance on the voices seen in training could be fruitful, as given a sufficiently large set of training voices selecting from this achieves the security goals of the system.

\subsection{Limitations of Proposed System}
\label{sec:limitations}
Through the experiments we evaluated in this paper, we demonstrated that improvements to identity generation methods, coupled with improvements to multi speaker TTS are needed to ensure that the space of producible voices is as wide as natural voices.

We expect that these improvements are likely to come with time, as computational power increases and increasingly advanced neural models for multi speaker TTS are developed.
Comparing progress from just a few years ago, when high quality TTS required large databases of audio for that individual, to now where a few seconds is all that is required to clone a voice, it does not seem unlikely that techniques will continue to improve in this area.

Further progress is also needed in audio intelligibility and understanding.
The best identity generation methods caused an increase in WER of approximately 15\% over the baseline and MOS scores were degraded from 4.23 to around 2.5 (2 is rated as poor and 3 as fair).
Again this is an area that has been rapidly improving in recent years with neural systems.

Performance improvements in this area are likely to come from both improvements in the STT and the TTS systems.

Furthermore in this work we only used the default Deepspeech system for STT, and adaption of this to specific tasks should improve the WER.
It's also worth noting that the dataset used for our WER analysis is from applying the system in an end to end manner on the LibriSpeech dataset, which was produced by people reading books aloud.
These books were from Project Gutenberg, and as such are generally over 95 years old~\cite{projectgutenberg},
meaning the text in them is likely to be different from the kind of speech used today.

For practical usage of the system there is also the difficulty of input lag.
By using VAD this lag can be minimised to the length of speaking without a pause, but it can not be eliminated.
Similarly there is also some lag introduced by the processing pipeline, and although each of the models operates with a real-time factor of below 1, we have not yet experimentally analysed the effects of input lag on the overall usability of the system, which we leave to future work.

Finally the system is limited due to the deterministic nature of the audio that it produces -- given identical text and identity embedding, the exact same audio will be produced.
In this work we only consider text-independent speaker recognition, but text-dependant is often used, where a specific phrase must be spoken.
However, with the AltVoice system the same phrase will always be identical, which may lead to systems preventing access due to detecting an overly similar audio sample i.e. a replay attack. 
Further work is needed to evaluate methods to alleviate this, such as by applying a small perturbation to the identity embedding to yield slightly different audio.

\subsection{Additional Secrets}
One of the principle attractions of voice based authentication is that the system user is no longer required to possess an explicit secret to access the service, with this secret instead being derived from the users voice.
Our proposed system reneges on this, and once again requires a user-known secret, in order to create the correct voice identity.
Whilst this may appear to be a step back, in a situation where the original voice trait, and thus the secret within it, has been compromised, it is necessary to introduce a new secret for voice authentication to continue to be used.
Unlike many biometrics, voice data is actively shared in many contexts, and as such the likelihood of a voice trait being exposed is high for most individuals.
AltVoice therefore allows such individuals to use remote voice based authentication systems with reduced fear of being the victim of an impersonation attack.

\subsection{Longer Term Implications}
The existence of a system such as AltVoice shifts the requirements of remote based speech systems.
In particular, the ability to change your voice will require remote speech systems to offer the ability to re-enroll if the voice for that remote system is compromised (\textit{revocability}, see ~\cite{Nandakumar2015}).
This would bring remote voice services in line with traditional text based password systems, where the password can later be changed by the user, but is unnecessary if methods to generate additional voices do not exist.

The compatibility of this system with existing remote systems is also likely to be a point of tension. 
In particular previous works have investigated being able to detect synthetically generated audio via liveness detection or spoofing detection e.g. ~\cite{Blue2018, Zhang2016,Ahmed2020, Wu2017, Kinnunen2017, todisco2019asvspoof}.
Such works include the ASVSpoof challenge, which has undergone several iterations, with systems achieving EERs below 1\% in the 2019 edition~\cite{todisco2019asvspoof}. As such audio produced using the proposed system may at this stage be caught by such a system.

However, for a user who has already had their voice trait stolen, and perhaps audio of them uttering specific keyphrases needed to access a given system, the ability to change their voice would be a benefit to them, as compared to relying on an adversary to not generate a convincing enough fake or to not have the correct phrase to play.
In particular the ASVSpoof challenge does not contain samples where an attacker has logical access to the system, and recordings of the original speaker, as would be the case with a remote replay attack (such as playing a sample directly down a phone).
Such samples should by definition be the same as regular speech down a phone, and thus would not be distinguishable from other samples.
AltVoice protects against an attacker conducting this attack, as the attacker will not be able to replay a sample of the victim's voice.

Thus a tradeoff arises, between synthetic voices being allowed and people having the benefits of this, such as no voice reuse and cancel-ability against systems attempting to prevent synthetic voice usage, leaving people potentially vulnerable to adversaries who can bypass this.
One possible solution is to enable spoofing detection checks by default, but allow users to disable these if they wish to use alternate voice systems.

\section{Related Work}
\subsection{Template Protection \& Cancellable Biometrics}

In biometric systems, recognition is typically performed using \textit{templates}, produced by feature extraction algorithms applied to data obtained from a sensor \cite{Nandakumar2015}.
The template(s) generated at enrolment are then compared with a template generated at authentication time to determine if the user is who they claim to be.

Template protection methods are applied to the templates stored in the system, in order to prevent the loss of a users templates to an attacker resulting in the ability to impersonate users.
This is done through the introduction of three specific properties: \textit{noninvertibility} - it should be computationally difficult to recover an individual's biometric template from a leaked protected one, \textit{revocability} - it should be possible to invalidate a template and as a consequence losing one template doesn't compromise future templates for an individual, and \textit{nonlinkability} - it is computationally hard to determine if two protected templates are derived from the same user~\cite{Nandakumar2015}.

In a cancelable biometric system~\cite{Ratha2007}, a fixed (per-user) distortion is applied to the biometric signal - either at a signal or feature level - and the remainder of the biometric system uses this distorted signal.
Many cancellable biometric systems have been proposed for a variety of modalities, such as fingerprint(\cite{Jin2004}), iris(\cite{JinyuZuo2009}) and face (\cite{Boult2006}) based systems.

Within cancellable biometric systems there are two main types of system~\cite{Rathgeb2011}.
\textit{Non-invertible transform} based systems are those that use a specific non-invertible transform to produce the cancellable effect.
Changes in parameters enable the updating of templates, and it does not matter if the parameters used are exposed.
In contrast to this, \textit{biometric salting} uses invertible transforms, the parameters for which are kept secret and are in many ways similar to using a salt in a normal password system.

Very few existing cancellable biometric systems have been introduced for voice biometrics.
A system has been proposed for voiceprints \cite{Xu2008}, which uses a group signature scheme to protect the voice with a key, allowing a voice to be used as a signature. 
In~\cite{Paulini2016} a technique for converting voice prints from a speaker recognition systems into a binary representation is proposed.
Mtibaa et al.~\cite{Mtibaa2018} propose a shuffling technique applied to a binary representation from a similar model to provide a cancellable voice biometric implementation at a system level.
These approaches have two drawbacks: i) the binary technique is developed on earlier systems and may not transfer to deep learning based methods, and ii) the protection only exists if the system creator decides to use it, requiring users to trust all voice services they use.

\subsection{Voice Privacy Systems}
Recently many approaches have been proposed to introduce various degrees of privacy into voice systems.
Qian et al.~\cite{Qian2018} introduced VoiceMask, a middle layer between speech services and the user to protect their voice's privacy, which uses vocal tract length normalization (VLTN) to warp the audio.
This prevents re-identification of the original speaker, but does not consider an adaptive attacker, and does not study the use of speaker recognition systems with the anonymised voice (or aim to support this).
Srivastava et al.~\cite{Srivastava2019}  perform an evaluation of this approach, as well as a VLTN based voice conversion approach~\cite{Sundermann2003} and a voice conversion approach using autoencoders~\cite{chou2019one}, under threat models with informed attackers, and find that they can reach similar performance to baseline systems if they have full access to the private voice system.
This demonstrates that speaker identifying information has been retained through the technique.%
This differs from our work, as we specifically ensure that no speaker identifying information can be retained after applying the technique, thus meaning an adversary gains no advantage from knowing the system details.

Recently the Voice Privacy Challenge~\cite{VoicePrivacyInitiative} has taken place to encourage the development of voice privacy systems.
The baseline system for the challenge was an implementation of the system proposed by Fang et al.~\cite{Fang2019}, which extracts X-vectors for speaker identity, as well as linguistic features, which are then re-synthesised with a new X-vector to create an anonymous voice.
The baseline technique proposed selecting a subset of X-vectors~\cite{Snyder2018} from a pool and averaging them to be the new X-vector, one of several strategies proposed for the system~\cite{srivastava2020design}.
Turner et al.~\cite{turner2020speaker, turner2022generating} proposed a method to generate more realistic x-vectors by training a Gaussian Mixture Model on a pool of speakers, which new X-vectors are then sampled from.
Mawalim et al.~\cite{Mawalim2020} also improved upon the baseline system by proposing a method for modifying the X-vector based on Singular Value Decomposition, and another based on reconstructing an X-vector from combining the output of two regression models.

The voice privacy challenge required participants to only remove the identity of the original speaker, and to maintain other speech characteristics, such as emotion and prosody.
Such an approach disqualifies AltVoice.
However, by requiring those components to be preserved there is an increased chance of information leakage that could be used by an attacker to determine the original speaker.
Our system differs from the solutions to the voice privacy challenge in that we anonymous the voice to the fullest extent possible, producing just text as our midpoint, leaving no room for any incidental transfer of identity information to occur.
However, we do make use of the X-vector selection techniques proposed in~\cite{srivastava2020design, turner2020speaker}, and adapt them for use with our embedding network for identity generation.

Han et al.~\cite{Han2020} propose a scheme for a privacy-preserving release of speech data, in which voice prints of the users in the dataset are protected.
This uses X-vectors as the voice print, and perturb this using an application of differential privacy to voices, termed voice-indistinguishability.
This approach is not usable by an end user, again requiring an end-user to trust the service provider to protect their voice appropriately. %

Abdullah et al.~\cite{abdullah2019hear} propose a method to attack widespread automatic speech recognition by mass surveillance systems by perturbing the audio at the word or phoneme levels using signal decomposition and reconstruction.
This prevents STT systems from inferring the textual content, and has some effect on disrupting speaker identification, however their is no perceptible change to the audio for humans, leaving users still vulnerable to more basic privacy invasions.
Similarly the attack does not try to move the voice to a specific user, and as such could not be used for any service that must be used repeatedly with the same identity, as AltVoice has been designed to do.

\section{Conclusion}
In this work we propose AltVoice, a system enabling users of remote voice system to both protect their privacy and use such systems without having to rely on the confidentiality of their possibly exposed voice trait.
The system reduces spoken audio to its textual content, before re-synthesising it with a newly generated identity embedding, resulting in audio that appears to be spoken by a different voice.

We examine several identity generation methods within the proposed system, evaluating them on the diversity of voice created, word error rates, and perceived audio quality.
We also examine them under two realistic threat models, for a privacy compromise attack and an authentication compromise attack.
Our results demonstrate that the new audio maintains its identity, and resists attacks under both threat models better than the status quo -- where a user's voice trait has already been exposed.

By outlining the system's trade-offs, we highlight how  improvements are needed for such a system.
In particular we find that re-using state-of-the-art TTS, STT and speaker recognition components in AltVoice leads to private voices with relatively high word error rates and limited perceived audio quality.
We expect many improvements to occur over time as the components mature and deep neural network research continues to provide performance gains.

Our system also raises new questions for the designers of remote voice systems, in particular as to how their systems may need rethinking to support the changing of voices over time, and how they can support systems such as the one proposed to facilitate better privacy for users, and better security for users with exposed voice traits.

\section*{Acknowledgments}
This work was supported by a grant from Mastercard and by the UK Engineering and Physical Sciences Research Council (EPSRC) [grant numbers EP/N509711/1, EP/P00881X/1].

\bibliographystyle{plain}
\bibliography{refs}

\begin{thebibliography}{10}

\bibitem{FirstDirectVoiceID}
{First Direct} phone banking - voice {ID} security.
\newblock https://www1.firstdirect.com/banking/ways-to-bank/telephone-banking/.

\bibitem{webrtc}
{WebRTC}.
\newblock https://webrtc.org/.

\bibitem{deepspeechrelease}
Deepspeech release 0.9.3 on {Github}.
\newblock https://github.com/mozilla/DeepSpeech/releases/tag/v0.9.3, 2021.

\bibitem{mozillacommonvoice}
Mozilla common voice dataset.
\newblock https://commonvoice.mozilla.org/en, 2021.

\bibitem{abdullah2019hear}
Hadi Abdullah, Muhammad~Sajidur Rahman, Washington Garcia, Logan Blue, Kevin
  Warren, Anurag~Swarnim Yadav, Tom Shrimpton, and Patrick Traynor.
\newblock Hear" no evil", see" kenansville": Efficient and transferable
  black-box attacks on speech recognition and voice identification systems.
\newblock {\em arXiv preprint arXiv:1910.05262}, 2019.

\bibitem{Ahmed2020}
Muhammad~Ejaz Ahmed, Il~Youp Kwak, Jun~Ho Huh, Iljoo Kim, Taekkyung Oh, and
  Hyoungshick Kim.
\newblock {Void: A fast and light voice liveness detection system}.
\newblock {\em Proceedings of the 29th USENIX Security Symposium},
  (1):2685--2702, 2020.

\bibitem{Blue2018}
Logan Blue, Luis Vargas, and Patrick Traynor.
\newblock {Hello, Is It Me You' re Looking For? Differentiating Between Human
  and Electronic Speakers for Voice Interface Security}.
\newblock {\em WiSec}, pages 123--133, 2018.

\bibitem{Boult2006}
T.~Boult.
\newblock {Robust distance measures for face-recognition supporting revocable
  biometric tokens}.
\newblock {\em FGR 2006: Proceedings of the 7th International Conference on
  Automatic Face and Gesture Recognition}, 2006:560--566, 2006.

\bibitem{chiu2018stateoftheart}
Chung-Cheng Chiu, Tara~N. Sainath, Yonghui Wu, Rohit Prabhavalkar, Patrick
  Nguyen, Zhifeng Chen, Anjuli Kannan, Ron~J. Weiss, Kanishka Rao, Ekaterina
  Gonina, Navdeep Jaitly, Bo~Li, Jan Chorowski, and Michiel Bacchiani.
\newblock State-of-the-art speech recognition with sequence-to-sequence models,
  2018.

\bibitem{chou2019one}
Ju-chieh Chou, Cheng-chieh Yeh, and Hung-yi Lee.
\newblock One-shot voice conversion by separating speaker and content
  representations with instance normalization.
\newblock {\em arXiv preprint arXiv:1904.05742}, 2019.

\bibitem{chung2018voxceleb2}
Joon~Son Chung, Arsha Nagrani, and Andrew Zisserman.
\newblock Voxceleb2: Deep speaker recognition.
\newblock In {\em Proc. Interspeech 2018}, pages 1086--1090, 2018.

\bibitem{cooper2020pretraining}
Erica Cooper, Xin Wang, Yi~Zhao, Yusuke Yasuda, and Junichi Yamagishi.
\newblock Pretraining strategies, waveform model choice, and acoustic
  configurations for multi-speaker end-to-end speech synthesis, 2020.

\bibitem{Fang2019}
Fuming Fang, Xin Wang, Junichi Yamagishi, Isao Echizen, Massimiliano Todisco,
  Nicholas Evans, and Jean-Francois Bonastre.
\newblock {Speaker Anonymization Using X-vector and Neural Waveform Models}.
\newblock pages 3--8, 2019.

\bibitem{NSASpying}
Electronic~Frontier Foundation.
\newblock {NSA} spying.
\newblock https://www.eff.org/nsa-spying, 2021.

\bibitem{projectgutenberg}
Project Gutenberg.
\newblock Frequently asked questions about project gutenberg.
\newblock https://www.gutenberg.org/help/faq.html.

\bibitem{Han2020}
Yaowei Han, Sheng Li, Yang Cao, Qiang Ma, and Masatoshi Yoshikawa.
\newblock {Voice-Indistinguishability: Protecting Voiceprint in
  Privacy-Preserving Speech Data Release}.
\newblock 2020.

\bibitem{Hannun2014}
Awni Hannun, Carl Case, Jared Casper, Bryan Catanzaro, Greg Diamos, Erich
  Elsen, Ryan Prenger, Sanjeev Satheesh, Shubho Sengupta, Adam Coates, and
  Andrew~Y. Ng.
\newblock {Deep Speech: Scaling up end-to-end speech recognition}.
\newblock pages 1--12, 2014.

\bibitem{hannun2014deep}
Awni Hannun, Carl Case, Jared Casper, Bryan Catanzaro, Greg Diamos, Erich
  Elsen, Ryan Prenger, Sanjeev Satheesh, Shubho Sengupta, Adam Coates, and
  Andrew~Y. Ng.
\newblock Deep speech: Scaling up end-to-end speech recognition, 2014.

\bibitem{HSBCVoiceID}
HSBC.
\newblock {Voice ID | HSBC UK}.
\newblock https://www.hsbc.co.uk/1/2/voice-id, 2018.

\bibitem{p808standard}
ITU-T.
\newblock Recommendation {P.808}: Subjective evaluation of speech quality with
  a crowdsourcing approach, 2018.

\bibitem{jia2019transfer}
Ye~Jia, Yu~Zhang, Ron~J. Weiss, Quan Wang, Jonathan Shen, Fei Ren, Zhifeng
  Chen, Patrick Nguyen, Ruoming Pang, Ignacio~Lopez Moreno, and Yonghui Wu.
\newblock Transfer learning from speaker verification to multispeaker
  text-to-speech synthesis, 2019.

\bibitem{Jin2004}
Andrew Teoh~Beng Jin, David Ngo~Chek Ling, and Alwyn Goh.
\newblock {Biohashing: Two factor authentication featuring fingerprint data and
  tokenised random number}.
\newblock {\em Pattern Recognition}, 37(11):2245--2255, 2004.

\bibitem{Kinnunen2017}
Tomi Kinnunen, Massimiliano Todisco, Nicholas Evans, Junichi Yamagishi, and
  Kong~Aik Lee.
\newblock {The ASVspoof 2017 Challenge : Assessing the Limits of Replay
  Spoofing Attack Detection National Institute of Informatics , Japan}.
\newblock (i):2--6, 2017.

\bibitem{Mawalim2020}
Candy~Olivia Mawalim, Kasorn Galajit, Jessada Karnjana, and Masashi Unoki.
\newblock {X-Vector Singular Value Modification and Statistical-Based
  Decomposition with Ensemble Regression Modeling for Speaker Anonymization
  System}.
\newblock In {\em Proc. Interspeech 2020}, pages 1703--1707, 2020.

\bibitem{Morris2004}
Andrew Morris, Viktoria Maier, and Phil Green.
\newblock From wer and ril to mer and wil: improved evaluation measures for
  connected speech recognition.
\newblock 01 2004.

\bibitem{Mtibaa2018}
Aymen Mtibaa, Dijana Petrovska-Delacr{\'{e}}taz, and Ahmed {Ben Hamida}.
\newblock {Cancelable speaker verification system based on binary Gaussian
  mixtures}.
\newblock {\em 2018 4th International Conference on Advanced Technologies for
  Signal and Image Processing, ATSIP 2018}, pages 1--6, 2018.

\bibitem{Mukhopadhyay2015}
Dibya Mukhopadhyay, Maliheh Shirvanian, and Nitesh Saxena.
\newblock {All your voices are belong to us: Stealing voices to fool humans and
  machines}.
\newblock {\em Lecture Notes in Computer Science (including subseries Lecture
  Notes in Artificial Intelligence and Lecture Notes in Bioinformatics)},
  9327:599--621, 2015.

\bibitem{nagrani2017voxceleb}
Arsha Nagrani, Joon~Son Chung, and Andrew Zisserman.
\newblock Voxceleb: A large-scale speaker identification dataset.
\newblock In {\em Proc. Interspeech 2017}, pages 2616--2620, 2017.

\bibitem{Nandakumar2015}
Karthik Nandakumar and Anil~K. Jain.
\newblock {Biometric template protection: Bridging the performance gap between
  theory and practice}.
\newblock {\em IEEE Signal Processing Magazine}, 32(5):88--100, 2015.

\bibitem{librispeech}
V.~{Panayotov}, G.~{Chen}, D.~{Povey}, and S.~{Khudanpur}.
\newblock Librispeech: An asr corpus based on public domain audio books.
\newblock In {\em 2015 IEEE International Conference on Acoustics, Speech and
  Signal Processing (ICASSP)}, pages 5206--5210, 2015.

\bibitem{Paulini2016}
M.~Paulini, C.~Rathgeb, A.~Nautsch, H.~Reichau, H.~Reininger, and C.~Busch.
\newblock {Multi-bit allocation: Preparing voice biometrics for template
  protection}.
\newblock {\em Odyssey 2016: Speaker and Language Recognition Workshop}, pages
  291--296, 2016.

\bibitem{Qian2018}
Jianwei Qian, Haohua Du, Jiahui Hou, Linlin Chen, Taeho Jung, and Xiang~Yang
  Li.
\newblock {Hidebehind: Enjoy voice input with voiceprint unclonability and
  anonymity}.
\newblock {\em SenSys 2018 - Proceedings of the 16th Conference on Embedded
  Networked Sensor Systems}, pages 82--94, 2018.

\bibitem{Ratha2007}
Nalini~K. Ratha, Sharat Chikkerur, Jonathan~H. Connell, and Ruud~M. Bolle.
\newblock {Generating cancelable fingerprint templates}.
\newblock {\em IEEE Transactions on Pattern Analysis and Machine Intelligence},
  29(4):561--572, 2007.

\bibitem{Rathgeb2011}
Christian Rathgeb and Andreas Uhl.
\newblock {A Survey on Biometric Cryptosystems}.
\newblock pages 1--25, 2011.

\bibitem{shen2018natural}
Jonathan Shen, Ruoming Pang, Ron~J Weiss, Mike Schuster, Navdeep Jaitly,
  Zongheng Yang, Zhifeng Chen, Yu~Zhang, Yuxuan Wang, Rj~Skerrv-Ryan, et~al.
\newblock Natural tts synthesis by conditioning wavenet on mel spectrogram
  predictions.
\newblock In {\em 2018 IEEE International Conference on Acoustics, Speech and
  Signal Processing (ICASSP)}, pages 4779--4783. IEEE, 2018.

\bibitem{Snyder2018}
David Snyder, Daniel Garcia-Romero, Gregory Sell, Daniel Povey, and Sanjeev
  Khudanpur.
\newblock {X-VECTORS : ROBUST DNN EMBEDDINGS FOR SPEAKER RECOGNITION David
  Snyder , Daniel Garcia-Romero , Gregory Sell , Daniel Povey , Sanjeev
  Khudanpur Center for Language and Speech Processing {\&} Human Language
  Technology Center of Excellence The Johns Hopkins Un}.
\newblock {\em Icassp2018}, pages 5329--5333, 2018.

\bibitem{srivastava2020design}
Brij Mohan~Lal Srivastava, Natalia Tomashenko, Xin Wang, Emmanuel Vincent,
  Junichi Yamagishi, Mohamed Maouche, Aurélien Bellet, and Marc Tommasi.
\newblock Design choices for x-vector based speaker anonymization, 2020.

\bibitem{Srivastava2019}
Brij Mohan~Lal Srivastava, Nathalie Vauquier, Md~Sahidullah, Aur{\'{e}}lien
  Bellet, Marc Tommasi, and Emmanuel Vincent.
\newblock {Evaluating Voice Conversion-based Privacy Protection against
  Informed Attackers}.
\newblock (825081):2--6, 2019.

\bibitem{Sundermann2003}
D.~{Sundermann} and H.~{Ney}.
\newblock {VTLN}-based voice conversion.
\newblock In {\em Proceedings of the 3rd IEEE International Symposium on Signal
  Processing and Information Technology (IEEE Cat. No.03EX795)}, pages
  556--559, 2003.

\bibitem{todisco2019asvspoof}
Massimiliano Todisco, Xin Wang, Ville Vestman, Md~Sahidullah, Hector Delgado,
  Andreas Nautsch, Junichi Yamagishi, Nicholas Evans, Tomi Kinnunen, and
  Kong~Aik Lee.
\newblock Asvspoof 2019: Future horizons in spoofed and fake audio detection,
  2019.

\bibitem{VoicePrivacyInitiative}
Natalia Tomashenko, Brij Mohan~Lal Srivastava, Xin Wang, Emmanuel Vincent,
  Andreas Nautsch, Junichi Yamagishi, Nicholas Evans, Jose Patino,
  Jean-Fran{\c{c}}ois Bonastre, Paul-Gauthier No{\'e}, and Massimiliano
  Todisco.
\newblock Introducing the {VoicePrivacy} initiative.
\newblock 2020.

\bibitem{turner2019attacking}
Henry Turner, Giulio Lovisotto, and Ivan Martinovic.
\newblock Attacking speaker recognition systems with phoneme morphing.
\newblock In {\em European Symposium on Research in Computer Security}, pages
  471--492. Springer, 2019.

\bibitem{turner2020speaker}
Henry Turner, Giulio Lovisotto, and Ivan Martinovic.
\newblock Speaker anonymization with distribution-preserving x-vector
  generation for the voiceprivacy challenge 2020, 2020.

\bibitem{turner2022generating}
Henry Turner, Giulio Lovisotto, and Ivan Martinovic.
\newblock Generating identities with mixture models for speaker anonymization.
\newblock {\em Computer Speech \& Language}, 72:101318, 2022.

\bibitem{Veaux2017CSTRVC}
C.~Veaux, J.~Yamagishi, and Kirsten Macdonald.
\newblock Cstr vctk corpus: English multi-speaker corpus for cstr voice cloning
  toolkit.
\newblock 2017.

\bibitem{wan2020generalized}
Li~Wan, Quan Wang, Alan Papir, and Ignacio~Lopez Moreno.
\newblock Generalized end-to-end loss for speaker verification, 2020.

\bibitem{Wu2017}
Zhizheng Wu, Junichi Yamagishi, Tomi Kinnunen, Cemal Hanil{\c{c}}i, Mohammed
  Sahidullah, Aleksandr Sizov, Nicholas Evans, Massimiliano Todisco, and
  H{\'{e}}ctor Delgado.
\newblock {ASVspoof: The automatic speaker verification spoofing and
  countermeasures challenge}.
\newblock {\em IEEE Journal on Selected Topics in Signal Processing},
  11(4):588--604, 2017.

\bibitem{xiong2017microsoft}
W.~Xiong, L.~Wu, F.~Alleva, J.~Droppo, X.~Huang, and A.~Stolcke.
\newblock The microsoft 2017 conversational speech recognition system, 2017.

\bibitem{Xu2008}
Wenhua Xu, Qianhua He, Yanxiong Li, and Tao Li.
\newblock {Cancelable voiceprint templates based on knowledge signatures}.
\newblock {\em Proceedings of the International Symposium on Electronic
  Commerce and Security, ISECS 2008}, (1):412--415, 2008.

\bibitem{yang2020multiband}
Geng Yang, Shan Yang, Kai Liu, Peng Fang, Wei Chen, and Lei Xie.
\newblock Multi-band melgan: Faster waveform generation for high-quality
  text-to-speech, 2020.

\bibitem{libritts}
Heiga Zen, Rob Clark, Ron~J. Weiss, Viet Dang, Ye~Jia, Yonghui Wu, Yu~Zhang,
  and Zhifeng Chen.
\newblock Libritts: A corpus derived from librispeech for text-to-speech.
\newblock In {\em Interspeech}, 2019.

\bibitem{Zhang2016}
Linghan Zhang, Sheng Tan, Jie Yang, and Yingying Chen.
\newblock {VoiceLive}.
\newblock pages 1080--1091, 2016.

\bibitem{JinyuZuo2009}
Jinyu Zuo, Nalini~K. Ratha, and Jonathan~H. Connell.
\newblock {Cancelable iris biometric}.
\newblock pages 1--4, 2009.

\end{thebibliography}

\end{document}